%% file: ms.tex
\documentclass[conference]{IEEEtran}
\usepackage{cite}
\usepackage{amsmath,amssymb,amsfonts}
\usepackage{algorithmic}
\usepackage{graphicx}
\usepackage{textcomp}
\usepackage{xcolor}
\def\BibTeX{{\rm B\kern-.05em{\sc i\kern-.025em b}\kern-.08em
    T\kern-.1667em\lower.7ex\hbox{E}\kern-.125emX}}

\usepackage{mathtools}
\usepackage{flushend}

\DeclareMathOperator*{\argmax}{arg\,max}

\usepackage{bm,afterpage,paralist,subfigure,array,caption,url}
\usepackage{tikz}
\usetikzlibrary{positioning, arrows.meta}
\usepackage{booktabs}
\usepackage{multicol}

\usepackage{array}
\newcolumntype{C}[1]{>{\centering\let\newline\\\arraybackslash\hspace{0pt}}m{#1}}

\newcommand{\citet}{\cite} 

\IEEEoverridecommandlockouts

\begin{document}

\title{On Linear Learning with Manycore Processors}

\author{\IEEEauthorblockN{Eliza Wszola}
\IEEEauthorblockA{
\textit{Department of Computer Science} \\
\textit{ETH Zurich}\\
Zurich, Switzerland \\
eliza.wszola@inf.ethz.ch}
\\
\IEEEauthorblockN{Martin Jaggi}
\IEEEauthorblockA{
\textit{School of Computer and Communication Sciences} \\
\textit{EPFL}\\
Lausanne, Switzerland \\
martin.jaggi@epfl.ch}
\and
\IEEEauthorblockN{Celestine Mendler-D\"unner$^\dagger$\thanks{$^\dagger$Work conducted while at IBM Research, Zurich.}}
\IEEEauthorblockA{
\textit{Department of Electrical Engineering and Computer Sciences} \\
\textit{UC Berkeley}\\
Berkeley, US \\
mendler@berkeley.edu}
\\
\IEEEauthorblockN{Markus P\"uschel}
\IEEEauthorblockA{
\textit{Department of Computer Science} \\
\textit{ETH Zurich}\\
Zurich, Switzerland \\
pueschel@inf.ethz.ch}
}

\maketitle

%
\begin{abstract}
A new generation of manycore processors is on the rise that offers dozens and more cores on a chip and, in a sense, fuses host processor and accelerator. In this paper we target the efficient training of generalized linear models on these machines. We propose a novel approach for achieving parallelism which we call Heterogeneous Tasks on Homogeneous Cores (HTHC). It divides the problem into multiple fundamentally different tasks, which themselves are parallelized. For evaluation, we design a detailed, architecture-cognizant implementation of our scheme on a recent 72-core Knights Landing processor that is adaptive to the cache, memory, and core structure. Our library efficiently supports dense and sparse datasets as well as 4-bit quantized data for further possible gains in performance. We show benchmarks for Lasso and SVM with different data sets against straightforward parallel implementations and prior software. In particular, for Lasso on dense data, we improve the state-of-the-art by an order of magnitude.
\end{abstract}


\renewcommand{\arraystretch}{1.2}

\newcommand{\eliza}{}\newcommand{\celestine}{}\newcommand{\martin}{}\newcommand{\markus}{}


\newcommand{\A}{\mathcal{A}}
\newcommand{\B}{\mathcal{B}}
\newcommand{\vv}{ {\bf v}}
\newcommand{\dv}{ {\bf d}}
\newcommand{\zv}{ {\bf z}}
\newcommand{\alphav}{ {\boldsymbol \alpha}}
\newcommand{\gap}{ {\text{gap}}}
\newcommand{\Real}{ {\mathbb{R}}}
\newcommand{\0}{ {\bf 0}}
\newcommand{\cP}{\mathcal{P}}
\newcommand{\nwv}{{\bf \hat{w}}}
\newcommand{\nvv}{{\bf \hat{v}}}

%
\maketitle

\section{Introduction}\label{sec:intro}

The evolution of mainstream computing systems has moved from the multicore to the manycore area. This means that a few dozen to even hundreds of cores are provided on a single chip, packaged with up to hundreds of gigabytes of memory at high bandwidth. Examples include Intel Xeon Phi (up to 72 cores), ARM ThunderX2 (64 cores), Qualcomm Centriq 2400 (48 cores), and of course GPUs (100s of cores). One declared target of the recent generation of manycores is machine learning. While much work has been devoted to efficient learning and inference of neural nets on GPUs, e.g.~\cite{TensorFlow, cuDNN}, other domains of machine learning and manycores have received less attention.

One exciting trend in manycore is the move from accelerators (like GPUs) to standalone manycore processors. These remove the burden of writing two types of code and enable easier integration with applications and legacy code. However, the efficient mapping of the required mathematics to manycores is a difficult task as compilers have inherent limitations to perform it given straightforward C (or worse, Java, Python, etc.) code, a problem that has been known already for the earlier, simpler multicore and single core systems~\cite{Moura:05}. Challenges include vector instruction sets, deep cache hierarchies, non-uniform memory architectures, and efficient parallelization.

The challenge we address in this paper is how to map machine learning workloads to manycore processors. We focus on recent standalone manycores and the important task of training generalized linear models used for regression, classification, and feature selection.
Our core contribution is to show that in contrast to prior approaches, which assign the same kind of subtask to each core, we can often achieve significantly better overall performance and adaptivity to the system resources, by distinguishing between two fundamentally different tasks. A subset of the cores will be assigned a task $\A$ that only reads the model parameters, while the other subset of cores will perform a task $\B$ that updates them. So in the manycore setting, while the cores are \emph{homogeneous}, we show that assigning them \emph{heterogeneous} tasks results in improved performance and use of compute, memory, and cache resources. The adaptivity of our approach is particularly crucial: the number and assignment of threads can be adapted to the computing platform and the problem at hand.

We make the following contributions:
\begin{enumerate}
\item We describe a novel scheme, consisting of two heterogeneous tasks, to train generalized linear models on homogeneous manycore systems. We call it Heterogeneous Tasks on Homogeneous Cores (HTHC). 

\item We provide a complete, performance-optimized implementation of HTHC on a 72-core Intel Xeon Phi processor. Our library supports both sparse and dense data as well as data quantized to 4 bits for further possible gains in performance. Our code is publicly available\footnote{https://github.com/ElizaWszola/HTHC}.

\item We present a model for choosing the best distribution of threads for each task with respect to the machine's memory system. We demonstrate that with explicit control over parallelism our approach provides an order of magnitude speedup over a straightforward OpenMP implementation.

\item We show benchmarks for Lasso and SVM with different data sets against straightforward parallel implementations and prior software. In particular, for Lasso on dense data, we improve the state-of-the-art by an order of magnitude.

\end{enumerate}

\section{Problem Statement \& Background}
\label{sec:background}

This section details the considered problem class, provides necessary background on asynchronous stochastic coordinate descent and coordinate selection, and introduces our target platform: the Intel Knights Landing (KNL) manycore processor.

\subsection{Problem specification}

We focus on the training of generalized linear models (GLMs). In mathematical terms this can be expressed by the following optimization task:
\begin{equation}
\min_{\alphav \in \mathbb{R}^n} \mathcal{F}({\boldsymbol \alpha}) := f(D{\boldsymbol \alpha}) +  \sum_{i\in [n]} g_i(\alpha_i),
\label{eq:obj}
\end{equation}
where $[n]=\{1,\dots,n\}$, $f$ and $g_i$ are convex functions, and  ${\boldsymbol \alpha\in  \mathbb{R}^n}$ is the model to be learned from the training data matrix $D\in \mathbb{R}^{d\times n}$ with columns ${\bf d}_1, \ldots, {\bf d}_n$. The function $f$ is assumed to be smooth. This general setup covers many widely applied machine learning models including logistic regression, support vector machines (SVM), and sparse models such as Lasso and elastic-net. 

\subsection{Coordinate descent}

A popular method for solving machine learning problems of the form \eqref{eq:obj} is stochastic (a.k.a. random) coordinate descent, where the separability of the term $g:= \sum_{i} g_i(\alpha_i)$ is crucial for its efficiency. Coordinate descent methods \cite{Wright:2015bn} form a group of algorithms which minimize $\mathcal{F}$ coordinate-wise, i.e., by performing the optimization across multiple iterations $i \in \{1, \dots, T\}$, 
where each iteration updates a single model parameter $\alpha_i$ using the corresponding data column ${\bf{d}}_i$, while the other parameters remain fixed. In this way, an optimization over a complex model with many variables can be split into a sequence of one-dimensional optimization problems.  Note that this approach can also be extended to batches where a subset of coordinates is updated at a time.  While the coordinate descent algorithm is by its nature sequential, it can be parallelized such that the coordinates are updated asynchronously in parallel by multiple threads. It can be shown that such a procedure maintains convergence guarantees if the delay between reading and updating the model vector is small enough~\cite{liu2015asynchronous}. This delay, which is a proxy for the staleness of the model information used to compute each update, depends on the data density and the number of threads working in parallel.

Typically, the coordinates to update are picked at random. However, to accelerate the coordinate decent procedure, we can assign  importance measures to individual coordinates. Different such measures exist and they depend either on the dataset, the current model parameters, or both. They can be used for the selection of important coordinates during the coordinate descent procedure, speeding up overall convergence,
either by deriving sampling probabilities (importance sampling, e.g.~\cite{Stich:2017tf, Perekrestenko:226287}), or by simply picking the parameters with the highest importance score (greedy approach, e.g.~\cite{You:2016:APG:3157382.3157621,karimireddy2019greedy}).

\subsection{Duality-gap based coordinate selection}\label{subsec:coordinate-selection}

A particular measure of coordinate-wise importance that we will adopt in this paper, is the coordinate-wise duality gap certificate proposed in~\citet{NIPS2017_7013}. The authors have shown that choosing model parameters to update based on their contribution to the duality gap provides faster convergence than random selection and classical importance sampling~\cite{pmlr-v37-zhaoa15}. 

To define the duality gap measure we denote the convex conjugates of $g_i$ by $g_i^\ast$, defined as $g_i^\ast(v) := \max_v v u-g_i(u)$. Then, the duality gap (see \cite{Boyd:2004uz}) of our objective \eqref{eq:obj} can be written as
\begin{align}
&\hspace{-0.3cm}\text{gap}({\boldsymbol \alpha};{\bf w}) = \sum_{i \in [n]} \text{gap}_i(\alpha_i; {\bf w}), \;\;\; \text{with}\notag \\
 &\hspace{-0.3cm}\text{gap}_i(\alpha_i;{\bf w}):= \alpha_i\langle{\bf w}, {\bf d}_i\rangle + g_i(\alpha_i) + g_i^\ast(-\langle{\bf w},{\bf d}_i \rangle),
\label{eq:gap}
\end{align}
where the model vector $\alphav, \bf w$ are related through the primal-dual mapping ${\bf w} := \nabla f(D {\boldsymbol \alpha})$. Importantly, knowing the parameters ${\alphav}$ and $ \bf w$, it is possible to calculate the duality gap values \eqref{eq:gap} for every $i\in[n]$ independently and thus in parallel.
In our implementation, we introduce the auxiliary vector $\vv = D \alphav$ from which ${\bf w}$ can be computed using a simple linear transformation for many problems of interest.

\subsection{The Knights Landing architecture}\label{subsec:knl}

Intel Knights Landing (KNL) is a manycore processor architecture used in the second generation Intel Xeon Phi devices, the first host processors, i.e., not external accelerators, offered in this line. It provides both high performance (with machine learning as one declared target) and x86 backwards compatibility. A KNL processor consists of 64--72\,cores with low base frequency (1.3--1.5\,GHz). 
KNL offers AVX-512, a vector instruction set for 512-bit data words, which allows parallel computation on 16\,single or 8\,double precision floats. It also supports vector FMA (fused multiply-add) instructions (e.g., $d = ab + c$) for further fine-grained parallelism. Each core can issue two such instructions per cycle, which yields a theoretical single precision peak performance of 64 floating point operations (flops) per cycle. Additionally, AVX-512 introduces gather-scatter intrinsics facilitating computations on sparse data formats.
The KNL cores are joined in pairs called tiles located on a 2D mesh. Each core has its own 32\,KB L1 cache and each tile has a 1\,MB L2 cache. The latter supports two reads and one write every two cycles. This bandwidth is shared between two cores. Each core can host up to four hardware threads.
KNL comes with two types of memory: up to 384\,GB of DRAM (6\,channels with an aggregate bandwidth of 80\,GB/s as measured with the STREAM benchmark~\cite{stream}) and 16\,GB of high-bandwidth MCDRAM (8 channels and up to 440\,GB/s respectively). The MCDRAM is configurable to work in one of three different modes:
\begin{inparaenum}[1)]
\item \textit{cache mode} in which it is used as L3 cache,
\item \textit{flat mode} in which it serves as a scratchpad, i.e., a software-controlled memory (in this mode, there is no L3 cache),
\item \textit{hybrid mode} in which part is working in cache mode and part in flat mode.
\end{inparaenum}
In this paper, we use a KNL with 72\,cores, 1.5\,GHz base frequency, and 192\,GB of DRAM in flat mode. The flat mode allows us to clearly separate the memory needed by the subtasks characterized in the next section.

\section{Method Description}

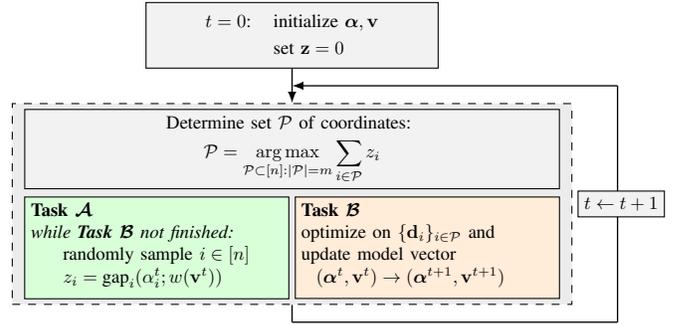
\begin{figure}[t]
\centering  
\input{duhl-diag}  
\caption{Visualization of our HTHC approach.}
\label{fig:duhl}
\end{figure}

Our scheme adopts an asynchronous block coordinate descent method where coordinate blocks are selected using the duality-gap as an importance measure as described in Section~\ref{subsec:coordinate-selection}. The workflow is illustrated in Figure~\ref{fig:duhl} and can be described as two tasks $\A$ and $\B$ running in parallel. Task~$\A$ is responsible for computing duality gap values $\text{gap}_i$ based on the current model $\alphav$ and the auxiliary vector $\vv$. These values are then stored in a vector $\zv\in\Real^n$ which we call gap memory. In parallel to task $\A$, task $\B$ performs updates on a subset of $m$ coordinates, which are selected based on their importance measure. For computing the updates on $\B$ we opt to use parallel asynchronous SCD (note that other importance sampling schemes or optimization algorithms could be applied to HTHC, as long as they allow $\B$ to operate on the selected columns $\dv_i$ in batches).
Since $\B$ operates only on a subset of data, it is typically faster than $\A$. Therefore, it is very likely that $\A$ is not able to update all coordinates of $\zv$ during a single execution of task $\B$ and some entries of the gap memory become stale as the algorithm proceeds. In practice, the algorithm works in epochs. In each epoch, $\B$  updates the batch of selected coordinates, where each coordinate is processed exactly once. At the same time, $\A$ randomly samples coordinates and computes $\text{gap}_i$  with the most recent (i.e., obtained in the previous epoch) parameters ${\alphav}, \vv$ and updates the respective coordinate $z_i$ of the gap memory.
As soon as the work of $\B$ is completed, it returns the updated $\alphav$ and $\vv$ to $\A$. $\A$ pauses its execution to select a new subset of coordinates to send to $\B$, based on the current state of the gap memory $\zv$. The robustness to staleness in the duality gap based coordinate selection scheme has been empirically shown in \cite{NIPS2017_7013}.

\subsection{Implementation challenges}

We identify the most computationally expensive parts of the proposed scheme.  Task $\A$ computes the coordinate-wise duality gaps \eqref{eq:gap}, which requires an inner product between ${\bf w}$, computed from $\vv$, and the respective data column~$\dv_i$:
\begin{equation}
z_i = h(\langle{\bf w}, {\bf d}_i\rangle, \alpha_i).
\label{eq:gapcompuation}
\end{equation}
$h$ is a scalar function defined by the learning model with negligible evaluation cost.

Task $\B$ performs coordinate descent on the selected subset of the data.  
Thus, in each iteration of the algorithm, a coordinate descent update on one entry of $\alphav$ is performed, i.e., $\alpha_i^+ = \alpha_i+\delta$. Also this update takes the form
\begin{equation}
\delta = \hat h (\langle{\bf w}, {\bf d}_i\rangle, \alpha_i) ,
\label{eq:updatecompuation}
\end{equation}
where~$\hat h$ is a scalar function. The optimal coordinate update has a closed-form solution~\cite{ss2013sdca,Wright:2015bn} for many applications of interest, and otherwise allows a simple gradient-step restricted to the coordinate~$i$.
With every update on $\alphav$ we also update $\vv$ accordingly: $\vv^+ = \vv + \delta \dv_i$, to keep these vectors consistent. 

The asynchronous implementation of SCD introduces two additional challenges: First, staleness of the model information $\vv$ used to compute updates might slow down convergence or even lead to divergence for a large number of parallel updates. Second, writing to shared data requires synchronization, and generates write-contention, which needs to be handled by appropriate locking.

\section{Implementation on Manycore Processors}
\label{sec:implementation}

The main contribution of this paper is to show that a scheme for learning GLMs based on multiple heterogeneous tasks is an efficient solution for implementation on a standalone, state-of-the-art manycore system such as KNL. As we will see, our approach is typically over an order of magnitude faster than simple C++ code with basic OpenMP directives. Due to the need for careful synchronization, locking and separation of resources, a straightforward implementation is not efficient in the manycore setting: a detailed calibration to the hardware resources and an associated implementation with detailed thread control is the key. 
In the following we will detail the challenges and key features for achieving an efficient implementation. 



\subsection{Parallelization of the workload}
\label{sec:parallelization}

Our implementation uses four levels of parallelism:
\begin{inparaenum}[1)]
\item $\A$ and $\B$ are executed in parallel.
\item $\A$ performs updates of $z_i$ in parallel and $\B$ performs parallel coordinate updates.
\item $\B$ uses multiple threads for each vector operation.
\item The main computations are vectorized using AVX-512.
\end{inparaenum}

\subsubsection{Allocation of resources to the tasks}
\label{sec:resourceallocation}

To map HTHC onto the KNL we divide compute and memory resources among the two tasks $\A$ and $\B$. We divide the compute resources by assigning separate sets of cores (in fact tiles for better data sharing) to each task. The respective number of cores is a parameter that makes our implementation adaptive to the problem and target platform. We use the threading library \textit{pthreads} for fine-grained control over thread affinity, synchronization, and locking over critical regions of the code. For more details we refer to Section \ref{sec:balancingAB}.
To split the memory resources between the two tasks we use the KNL in flat mode where the MCDRAM memory serves as a scratchpad. This setting is particularly suited for our scheme because we can allocate the data for $\A$ to DRAM and the data for $\B$ to MCDRAM. As a consequence, saturating the memory bandwidth by one task will not stall the other. This approach has another advantage. While the large datasets could not fit entirely into MCDRAM, $\B$ can be configured to work only with a subset of data small enough to be allocated there.

\subsubsection{Parallelization of the individual tasks}

For~$\A$, we use only one thread for every update of a single $z_i$ due to the high risk of deadlocks when computations on $\B$ are finished and $\A$ receives a signal to stop. 
The number $T_\A$ of threads used for~$\A$ is a parameter used for adaptation.

In contrast to $\A$, $\B$ performs $T_\B$ updates in parallel and also parallelizes the inner product computation of each update across $V_\B$ threads. Thus, the total number of threads used by $\B$ is $T_\B\cdot V_\B$. Both are parameters in our implementation. When $V_\B$ threads are used per update, $\vv$ and the corresponding $\dv_i$ are split into equal chunks. 

A simple model can be used to determine a good choice for $V_\B$ as explained next. The performance of both the inner product and the $\vv$ update is limited by the memory bandwidth. For this reason, it is desirable that $\vv$, which is reused, stays in cache. To achieve this, the cache has to hold $\vv$ and two columns $\dv_i$, $\dv_j$. Since $\vv$ and $\dv_i$ have the same length, this means the chunk size should be about a third of the cache size, i.e., about 87,000 single precision numbers for the L2 caches in KNL. Optimizing with the same reasoning for the 32KB L1 cache would yield a length of $\vv$ below 4096 elements. Such short vectors would not benefit from parallelism due to issues discussed later. Thus, we do not consider this setup applicable to the L1 caches. The best choice for $T_\B$ is influenced by several factors as will be discussed in Section \ref{sec:balancingAB}. 


\subsubsection{Vectorization with AVX-512}

We implemented both the scalar product (executed on both $\A$ and~$\B$) and the incrementation of ${\vv}$ (performed on $\B$) using AVX-512 FMA intrinsics with multiple accumulators for better instruction-level parallelism. The peak single core performance of KNL is 64 flops/cycle, but in the scalar product, each FMA requires two loads from L2 cache, reducing the peak to 16. In practice, our entire coordinate update achieves about 7.2 flops/cycle, about three times faster than without AVX.

%

\subsection{Synchronization}

Task $\A$ does not write to shared variables and thus requires no synchronization between threads.
In contrast, the updates on $\B$ are performed with multiple threads per vector as explained above. For the updates in Equation~\eqref{eq:updatecompuation}, three barriers are required to separate the resetting of the shared result from the 
scalar product and the computation of $\hat{h}$ based on the new shared result.

For the implementation we use pthreads which provides synchronization mechanisms with mutexes and thread barriers. Since barriers are relatively expensive, we replace them with a mechanism based on integer counters protected by mutexes similar to~\cite{SpiralBar}.



In addition to synchronization per thread, we need to coordinate running and stopping the tasks at the beginning and the end of each epoch $t$ (see Fig.~\ref{fig:duhl}). To avoid the overhead of creating and destroying threads, we use a thread pool with a constant number of threads for $\A$ and $\B$. To synchronize, we use another counter-based barrier scheme similar to the one described above. 

\subsection{Atomic operations}\label{sec:atomic}

We enforce atomic updates to the shared vector $\vv$ to preserve the primal-dual relationship between ${\bf w}$ and ${\boldsymbol \alpha}$ and thus maintain the convergence guarantees of asynchronous SCD derived by Hsieh et al.~\citet{hsieh2015passcode}. 
The pthreads library does not provide atomic operations, but the mutexes can be used to lock chosen variables. To avoid overhead, we use medium-grained locks for chunks of 1024 vector elements.

\subsection{Sparse representation}

To efficiently support also sparse datasets, we use a special data structure for $D$ akin to the CSC (compressed sparse-column) format, while $\vv$ and $\alphav$ remain in dense format. $D$ is represented as an array of structures containing pointers, one for each column. Each column contains only the nonzero elements, encoded as (index, value) pairs. $\B$ stores its own data columns $\{\dv_i\}_{i \in \cP}$ in a similar way, with the columns additionally split into chunks of a fixed length, implemented as linked lists. This way, efficient movement of columns of variable length between $\A$ and $\B$ into preallocated space is possible, accommodating possible large differences in column length. The minimal chunk size of 32 enables the use of multiple AVX-512 accumulators, but the optimal size depends on the density of $D$. We use locking as described in the previous section. Since the locks are fixed to equal intervals of the dense vector $\vv$, the number of operations performed under a given lock depends on the density of the corresponding interval of $\dv_i$ and 1024 might no longer be an efficient choice of the lock size and vary on each dataset.
%
%
Initially, $\B$ allocates a number of empty chunks determined by the $m$ densest columns $\dv_i$ in $D$, and places the chunks on a stack. When $\A$ copies data to $\B$, the pointers to chunks are obtained from the stack and rearranged into the linked lists, long enough to accommodate the new set of columns processed by $\B$. Next, the data of $\{\dv_i\}_{i \in \cP}$ is copied to the corresponding lists. At the same time, the pointers to the chunks linked by the lists corresponding to the columns which are swapped out of $\B$ are returned to the stack. With this representation, we observe fastest runtime when one thread is used per vector: in most cases, the sparse vectors are shorter than 130,000 elements.

\subsection{Quantized representation}

Stochastic quantization to reduced bit width reduces data size while still maintaining performance for many iterative optimization and machine learning algorithms (e.g.,~\cite{zhang2017zipml}). To investigate and provide potential benefits, we extend HTHC with support for 4-bit quantization using an adaptation of the Clover library~\cite{stojanov2018fast}, which provides efficient quantized low-level routines including the scalar product.
We find that 4-bit precision is enough to represent the data matrix $D$, without significantly sacrificing model accuracy.
For $\vv$ and $\alphav$, low precision results in excessive error accumulation; thus we leave those at 32-bit floating point.
The overall benefit is reduced data movement and memory consumption at the overhead of packing and unpacking 4-bit data for computation. We show runtime results in Section~\ref{sec:exp}.


\subsection{Balancing compute resources}
\label{sec:balancingAB}

A major challenge posed by the implementation of HTHC is how to balance the computing resources across the different levels of parallelism as discussed in Section~\ref{sec:parallelization}. The configuration of HTHC is parameterized by $T_\A$, $T_\B$, and $V_\B$, and can be adjusted to the hardware and problem at hand. We identified two important factors that impact the optimal setting:

\paragraph{Balanced execution speed} If $\B$ works significantly faster than $\A$, the latter executes only few $z_i$ updates. As a consequence most coordinate importance values become stale, and convergence suffers. This effect has empirically been investigated in \cite{NIPS2017_7013}, which showed that satisfactory convergence requires about 15\% or more of the $z_i$ being updated in each epoch. We will discuss this further in Section~\ref{sec:exp}.
On the other hand, if $\B$ is too slow, the runtime suffers. Hence, the efficiency of the implementation is a crucial factor that impacts the best configuration. 


\paragraph{Cache coherence} The parallelization of the gap memory updates on $\A$ across a large number of threads can lead to DRAM bandwidth saturation. Additionally, more threads mean higher traffic on the mesh, which can impact the execution speed of $\B$.
For fast convergence, the threads must be assigned so that $\A$ performs a sufficient fraction of $z_i$ updates in each epoch. Our results will confirm $\tilde r = 15\%$ of the columns of $D$ updated by $\A$ as a safe choice.


\paragraph{Performance model} Let us consider dense data. Recall that we operate on the data matrix $D\in \mathbb{R}^{d\times n}$, where each of the $n$ coordinates corresponds to a column represented by vector $\dv_i$ of length $d$, and that $\B$ processes $m$ coordinates per epoch. Let $t_{\mathcal{I},d}(\dots)$ denote the time of a single coordinate update on task $\mathcal{I} \in \{\A, \B\}$ with vector length $d$. This function is not trivial to derive, due to relatively poor scalability of the operations used and the dependence on memory and synchronization speed. Thus, we precompute the values for different thread setups and $d$ during installation and store them in a table. Using this table, we then use the following model to obtain the thread counts:
\begin{equation*}
\begin{aligned}
 \min_{m, T_\A, T_\B, V_\B} m \cdot t_{\B,d}(T_\B, V_\B) \;\; \text{s.t.} \; \; \; \frac{m \cdot t_{\B,d}(T_\B, V_\B)}{t_{\A,d}(T_\A)} \geq \tilde{r} \cdot n
\end{aligned}
\label{eq:best_params}
\end{equation*}



\section{Experimental Results}\label{sec:exp}

We perform two sets of experiments. In the first set, we profile HTHC on dense synthetic data with the aim to understand and illustrate how the different implementation parameters impact its performance. In the second set, we benchmark HTHC on KNL on real-world data from small to very large. We compare against a number of more straightforward variants of implementing the same algorithm including standard C++ code with OpenMP directives, and against prior software where available.

All experiments are run on a KNL in flat mode as described in Section~\ref{subsec:knl}. We compile our code with the Intel Compiler Collection and flags -std=c++11 -pthread -lmemkind -lnuma -O2 -xCOMMON-AVX512 -qopenmp. In all experiments, we use at most one thread per core and single precision.

\begin{figure}[t!]
\centering    
\includegraphics[width=0.7\columnwidth]{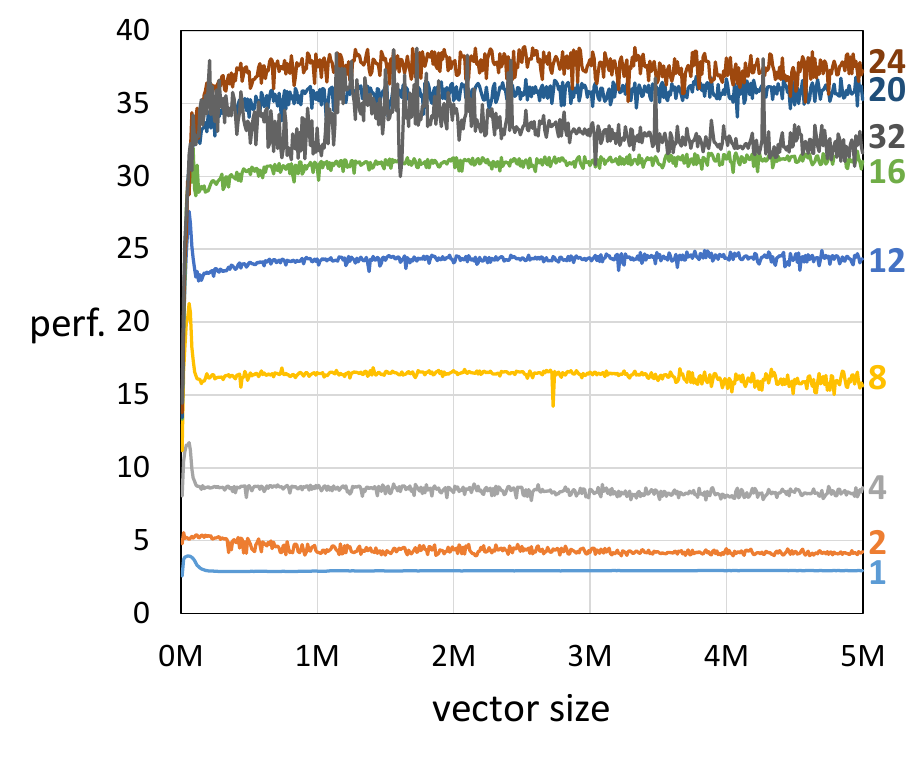}
\caption{Performance (in flops/cycle) of synthetic $\A$ operations. Different labels represent different values of $T_\A$.}
\label{fig:a-perf}
\end{figure}

\begin{figure*}[t]
\centering    
 \begin{minipage}[t]{\textwidth}
\subfigure[$T_\B=1$]{\label{fig:b-perf-1}\includegraphics[width=0.245\textwidth]{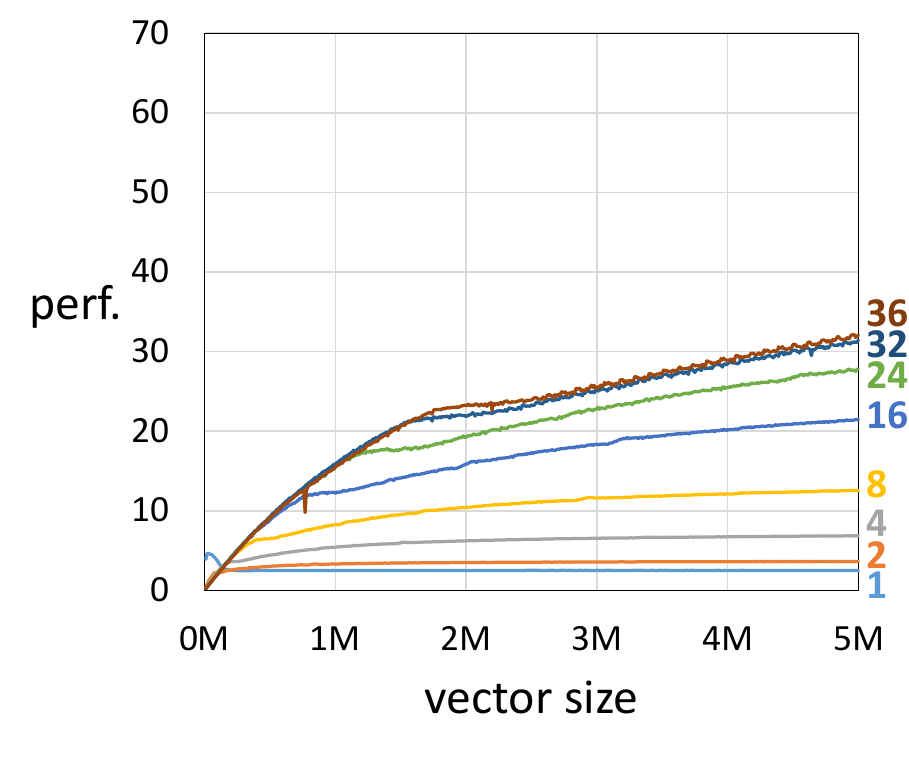}}
\subfigure[$T_\B=4$]{\label{fig:b-perf-4}\includegraphics[width=0.245\textwidth]{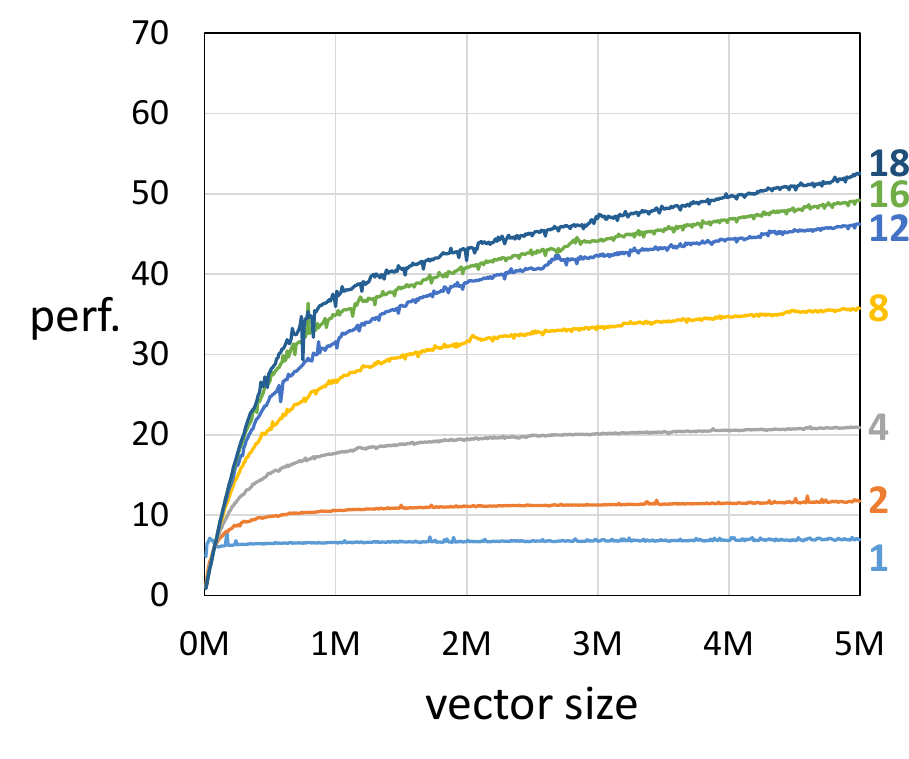}}
\subfigure[$T_\B=8$]{\label{fig:b-perf-8}\includegraphics[width=0.245\textwidth]{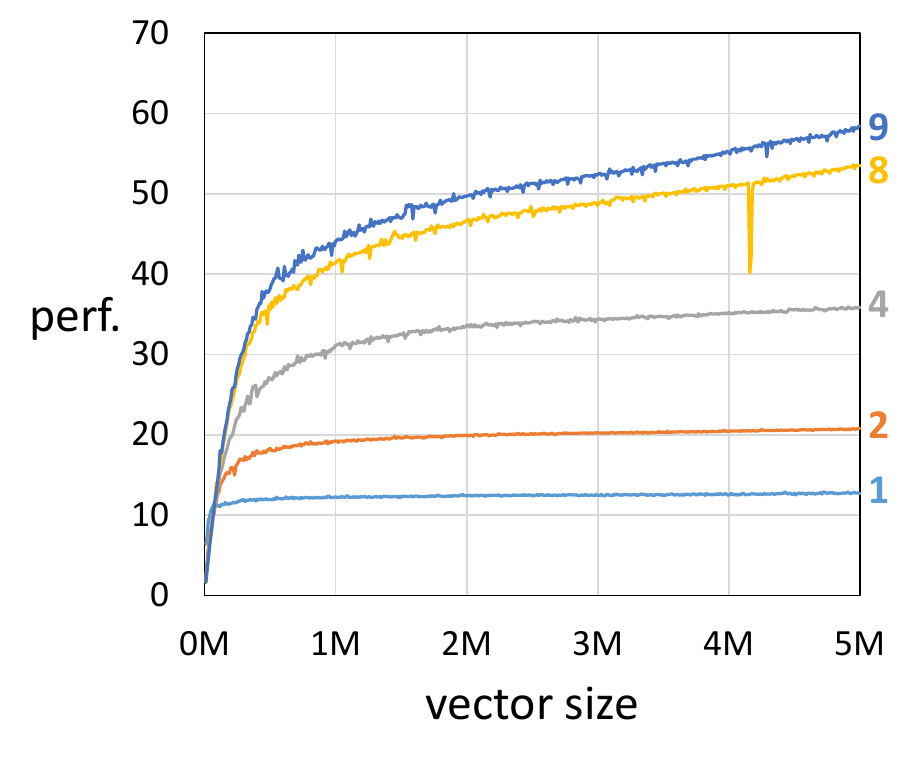}}
\subfigure[$T_\B=16$]{\label{fig:b-perf-16}\includegraphics[width=0.245\textwidth]{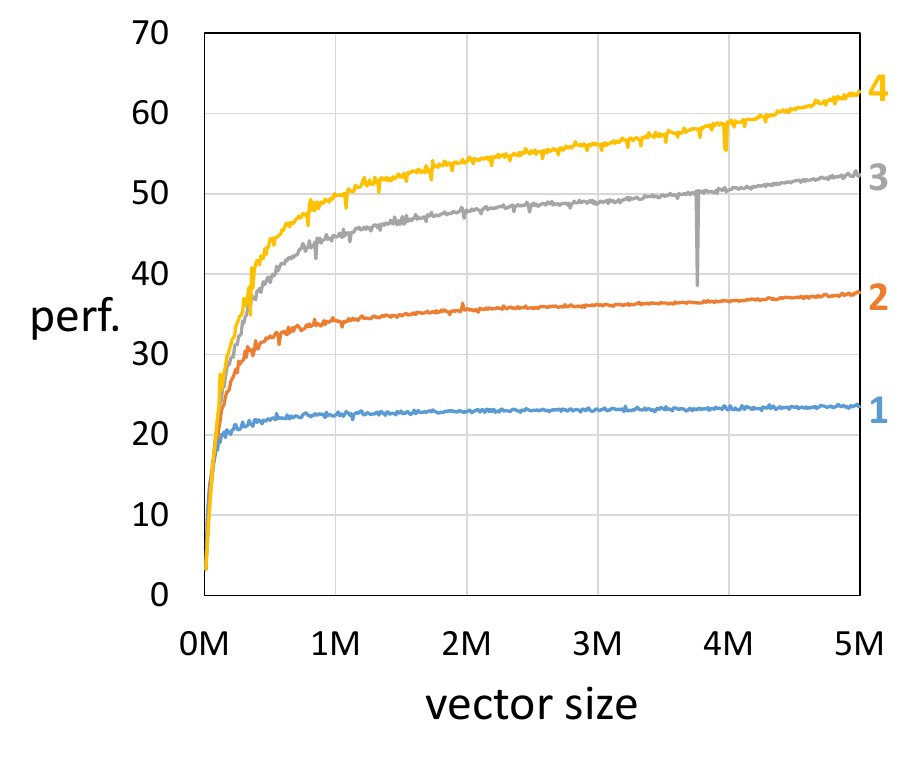}}
\caption{Performance (in flops/cycle) of operations of task $\B$ for different numbers $T_\B$ of  parallel updates. Different curves represent different values of $V_\B$.}
\label{fig:b-perf}
\end{minipage}
\end{figure*}

\subsection{Algorithm profiling}

To simulate different values of $t_{\mathcal{I},d}$ for different vector sizes $d$ (Section \ref{sec:balancingAB}), we imitate the expensive operations of the tasks $\A$ and $\B$ on dense synthetic data. The code measures the overall time and performance for different vector sizes and thread numbers. The operations involve the data matrix $D$ of size $n \times d$ and the shared vector ${\bf v}$, with threading and synchronization implemented as described in Section \ref{sec:implementation}. In the following we will illustrate results for varying data size ($n = 600$ and $d$ ranging from 10,000 to 5,000,000). 

To analyze the impact of the parameter $T_\A$ on the performance of task $\A$, we allocate both data structures to DRAM and measure performance for $T_\A$ ranging from 1 to 72. The results are presented in Fig.~\ref{fig:a-perf}. We observe that above 20 parallel updates, the performance does not increase significantly and above 24 it even begins to decrease and fluctuate due to the saturation of DRAM bandwidth. For this reason, we use at most 24 threads for $\A$. Note that more threads could be added if $\A$ operated on MCDRAM, however, many datasets are too large to fit there. Moreover, such a setup does not allow a clean separation of the resources of $\A$ and $\B$. This means that memory accesses would cause interference between the two, leading to slowdown.

To analyze the impact of the parameters $V_\B$ and $T_\B$ on the performance of task $\B$, we allocate $D$ and ${\bf v}$ to MCDRAM. Fig.~\ref{fig:b-perf} illustrates the impact of $V_\B$ and shows results for $T_\B = \{1, 4, 8, 16\}$. The few outliers in plots drawn for larger numbers of threads are caused by background processes which stall the execution of the program on particular cores. We note that below $d = 130,000$ it is best to use one thread per vector, independent of the number of parallel updates. For larger vectors, the best strategy is to use as many threads per vector as possible. We observe that for the vector lengths considered, higher performance is obtained with more parallel updates rather than with more threads per vector. This can be attributed to the overhead of synchronization when multiple threads work with the same vector.

\begin{figure}[t!]
\centering    
\includegraphics[width=0.7\columnwidth]{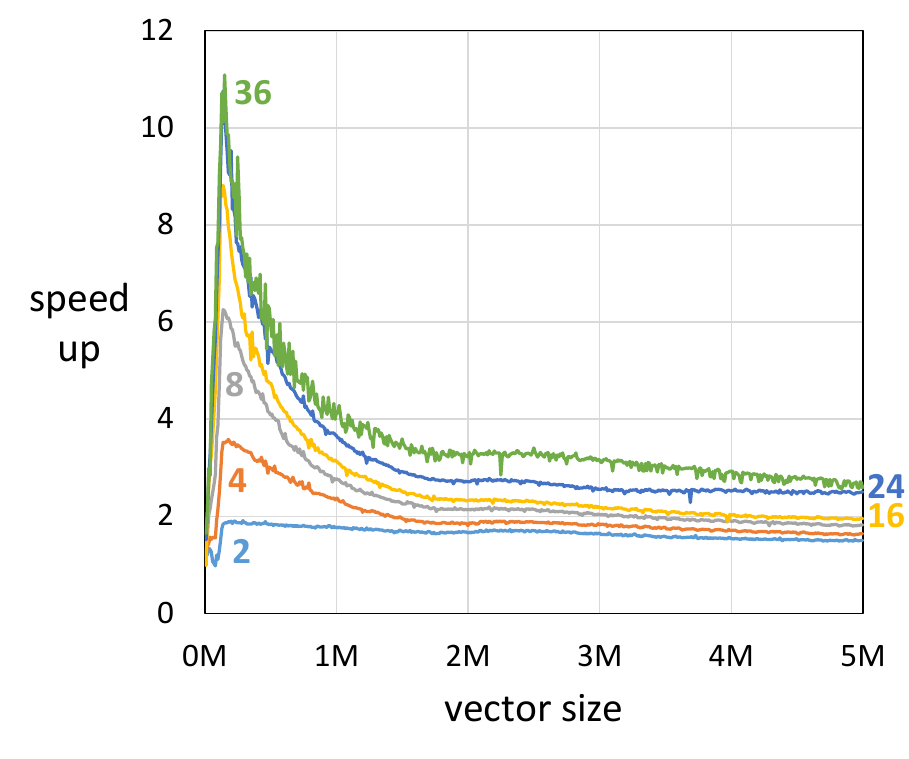}
\caption{Speedup of runs with different number $T_\B$ of parallel updates over runs with a single update on $\B$.}
\label{fig:b-speedups}
\end{figure}

Fig.~\ref{fig:b-speedups} shows the speedup of isolated $\B$ runs with different values of $T_\B$ over a run with $T_\B=1$. For each value of $T_\B$, we plot results for the runs with the best corresponding $V_\B$. We observe that the algorithm used by $\B$ does not scale well. This is due to many synchronization points during updates. Profiling with Intel VTune shows that while the bandwidth of L2 caches is a bottleneck on each tile, the saturation of MCDRAM bandwidth remains low. For this reason, we benefit from the flat mode, since it keeps MCDRAM as a separate allocation space.
The raw update speed of $\B$, contrary to the convergence of the complete scheme, is not affected by too many parallel updates of $\vv$. In practice, the optimal value for $T_\B$ is rarely the maximum, as we will see in the following experiments.

\begin{table}[t!]
\footnotesize
  \caption{Data sets used in the experiments}
  \label{tab:datasets}
  \centering
  \begin{tabular}{@{}lrrrr@{}}
    \toprule
    Dataset & Samples & Features & Representation & Approx. Size \\
    \midrule
    Epsilon~\cite{pascalChallenge} & 400,000 & 2,000 & Dense & 3.2 GB \\
    DvsC~\cite{heinze2016dual} & 40,002 & 200,704 & Dense & 32.1 GB\\
    News20~\cite{LIBSVMBinary} & 19,996 & 1,355,191 & Sparse & 0.07 GB\\
    Criteo~\cite{criteoChallenge} & 45,840,617 & 1,000,000 & Sparse & 14.4 GB\\
    \bottomrule
  \end{tabular}
\end{table}

\subsection{Performance evaluation}

The second series of experiments compares the performance of HTHC to several reference schemes of the two selected linear models across three data sets of different size.
We consider Lasso and SVM on the two dense and two sparse data sets in Table~\ref{tab:datasets}. Dogs vs.~Cats (abbrieviated as DvsC in our tables) features were extracted as in~\cite{heinze2016dual} and the number of samples was doubled. The same pre-processing was used in~\cite{NIPS2017_7013}.
The regularization parameter $\lambda$ was obtained to provide a support size of  12\% for Lasso on Epsilon and Dogs vs.~Cats, and using cross validation in all other cases.

\subsubsection{Comparison to our baselines}

In the following we will denote HTHC as $\A + \B$ emphasizing that it runs two tasks, $\A$ and $\B$. As detailed in Section~\ref{sec:resourceallocation}, HTHC allocates the data for $\A$ to DRAM and the data for $\B$ to MCDRAM. For each experiment, except those on the Criteo dataset, we used exhaustive search to find the best parameter settings, i.e., percentage of data updated by $\B$ per epoch $\%_\B$, and the thread settings $T_\A, T_\B, V_\B$ described in Section~\ref{sec:implementation}. The obtained parameters presented in Tables~\ref{tab:lasso-params},~\ref{tab:svm-params} roughly correspond to the analysis in Section~\ref{sec:balancingAB}.
We compare HTHC against four reference implementations:

\begin{table}[t!]
\footnotesize
  \caption{Best parameters found for Lasso.}
  \label{tab:lasso-params}
  \centering
  \begin{tabular}{@{}lrrrrrr@{} m{8pt} @{}rrr@{}}
    \toprule
    Data set & $\lambda$ & \multicolumn{5}{c}{settings for $\A+\B$} && \multicolumn{3}{@{}c@{}}{settings for ST} \\
    \cmidrule{3-7} \cmidrule{9-11}
     &  &  $\%_\B$ & $T_\A$ & $T_\B$ & $V_\B$ & $T_{\text{total}}$ && $T_\B$ & $V_\B$ & $T_{\text{total}}$ \\
    \midrule
    Epsilon & 3e-4 & 8\% & 12 & 8 & 6 & 60 && 8 & 9 & 72 \\
    DvsC & 2.5e-3 & 2\% & 16 & 14 & 1 & 30 && 20 & 1 & 20 \\
    News20 & 1e-4 & 2\% & 24 & 12 & 1 & 36 && 56 & 1 & 56 \\
    Criteo & 1e-6 & 0.1\% & 8 & 64 & 1 & 72 && 72 & 1 & 72 \\
    \bottomrule
  \end{tabular}

  \end{table}
  
  \begin{table}[t!]
  \footnotesize
  \caption{Best parameters found for SVM.}
  \label{tab:svm-params}
  \centering
  \begin{tabular}{@{}lrrrrrr@{} m{8pt} @{}rrr@{}}
    \toprule
    Data set & $\lambda$ & \multicolumn{5}{c}{settings for $\A+\B$} && \multicolumn{3}{c}{settings for ST} \\
    \cmidrule{3-7} \cmidrule{9-11}
     &  &  $\%_\B$ & $T_\A$ & $T_\B$ & $V_\B$ & $T_{\text{total}}$ && $T_\B$ & $V_\B$ & $T_{\text{total}}$ \\
    \midrule
    Epsilon & 1e-4 & 4\% & 16 & 2 & 1 & 18 && 2 & 1 & 2 \\
    DvsC & 1e-4 & 7\% & 8 & 6 & 10 & 68 && 36 & 2 & 72\\
    News20 & 1e-5 & 49\% & 12 & 56 & 1 & 72 && 72 & 1 & 72 \\
    Criteo & 1e-6 & 1\% & 4 & 68 & 1 & 72 && 72 & 1 & 72 \\
    \bottomrule
  \end{tabular}
\end{table}

\begin{enumerate}
\item ST (single task): We consider a parallel, but homogeneous single task implementation, which allocates the data matrix $D$ to DRAM and the remaining data to MCDRAM. It performs randomized asynchronous SCD. We used the same low-level optimizations in ST as in task $\B$ of HTHC but without duality-gap-based coordinate selection. Instead, in each epoch we update $\vv, \alphav$ (allocated to MCDRAM) for all coordinates of $D$. 
Again, we run a search for the best parameters. These are shown in Tables~\ref{tab:lasso-params} and~\ref{tab:svm-params}.
\item ST ($\A + \B$): Like ST but run with the best setting of $T_\B$ and $V_\B$ for $\A + \B$.
\item OMP: A straightforward implementation of $\A + \B$: standard looped C code using the OpenMP directives \texttt{simd reduction} and \texttt{parallel for} for parallelization with the thread counts $T_\A$, $T_\B$ and $V_\B$. To synchronize the updates of $\vv$, we use the directive \texttt{atomic}.
\item OMP WILD is as OMP, but without the \texttt{atomic} directive.
\end{enumerate}
We perform OMP runs only for the dense representations. For the large Criteo dataset, we consider only $\A+\B$ and ST due to the long time to run all experiments.

\begin{figure*}[th!]
\captionsetup[subfigure]{labelformat=empty}
\centering    
 \begin{minipage}[t]{\textwidth}
  \begin{tabular}{C{.39\textwidth} C{.22\textwidth} C{.4\textwidth}}
  Lasso, Epsilon & Lasso, Epsilon & SVM, Epsilon \\
  \end{tabular}
\subfigure{ \label{fig:lasso-eps-sub}\includegraphics[width=0.33\columnwidth]{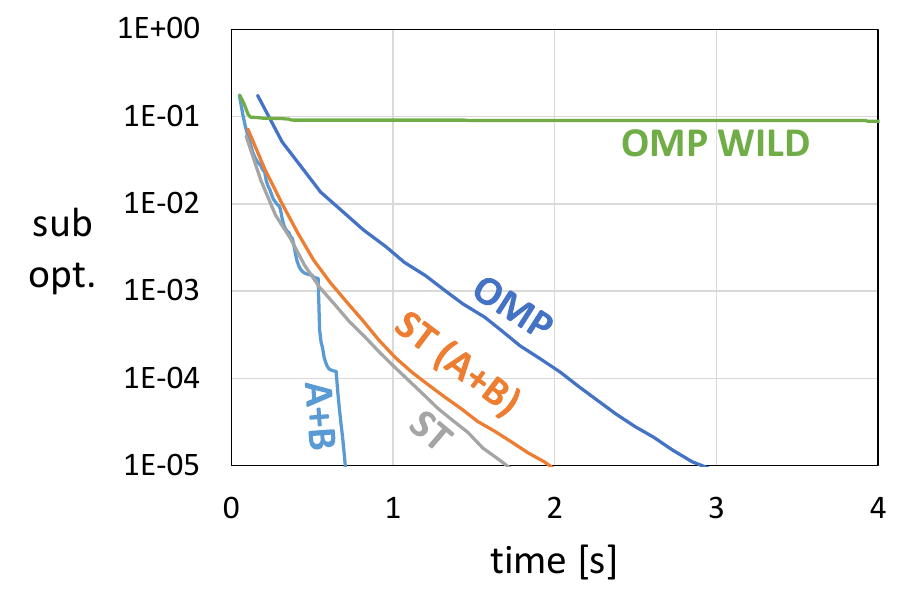}}
\subfigure{\label{fig:lasso-eps}\includegraphics[width=0.33\columnwidth]{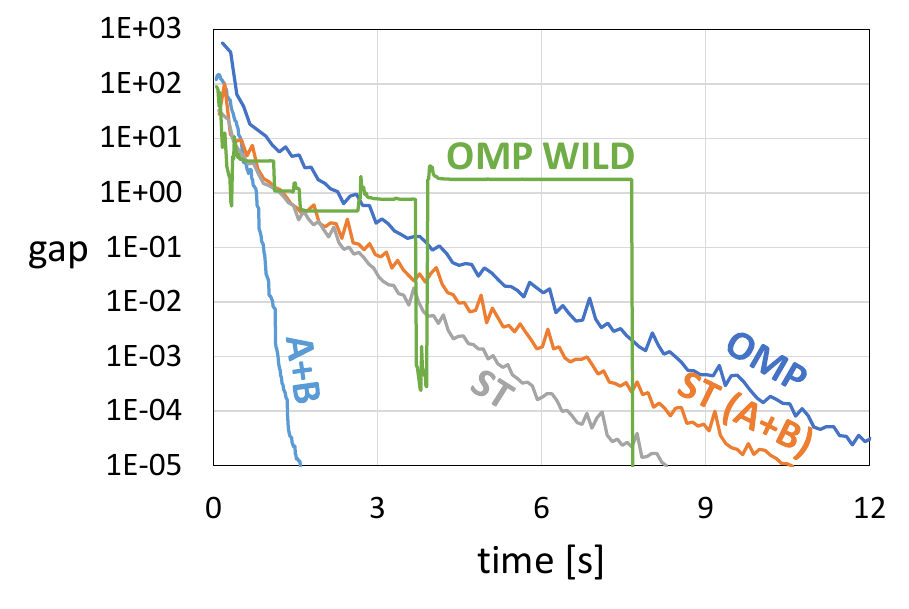}}
\subfigure{\label{fig:svm-eps}\includegraphics[width=0.33\columnwidth]{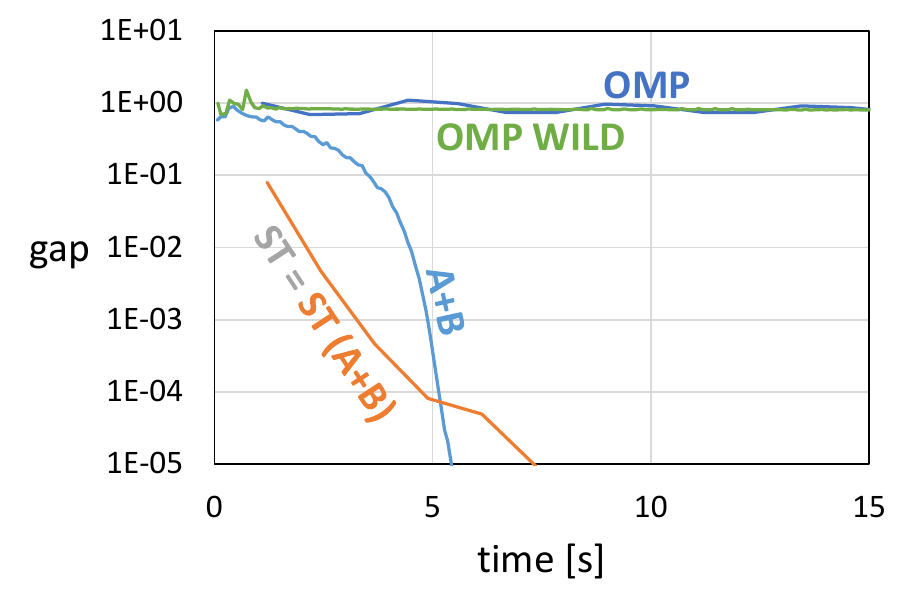}}
\end{minipage}
 \begin{minipage}[t]{\textwidth}
   \begin{tabular}{C{.39\textwidth} C{.22\textwidth} C{.4\textwidth}}
  Lasso, Dogs vs. Cats & Lasso, Dogs vs. Cats & SVM, Dogs vs. Cats \\
  \end{tabular}
\subfigure{\label{fig:lasso-dvc-sub}\includegraphics[width=0.33\columnwidth]{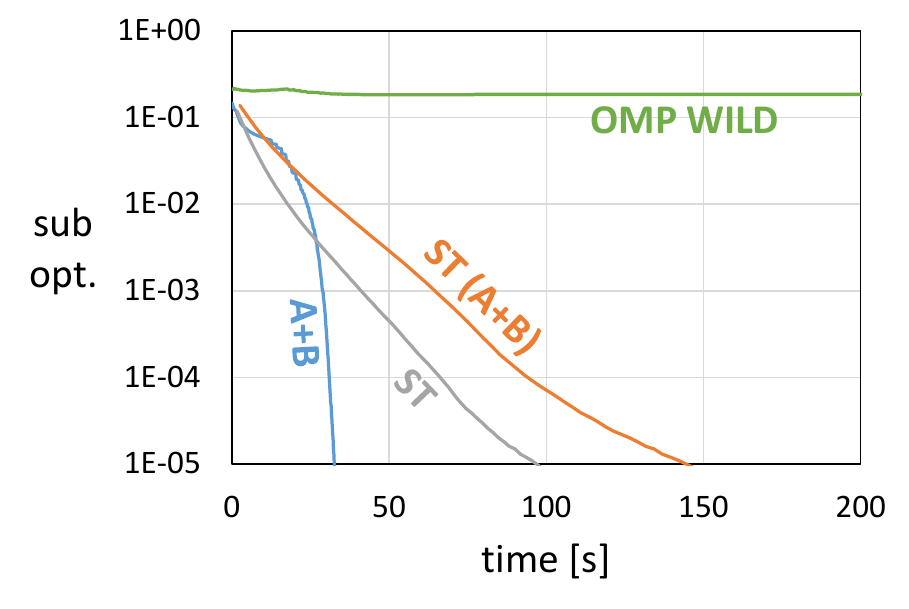}}
\subfigure{\label{fig:lasso-dvc}\includegraphics[width=0.33\columnwidth]{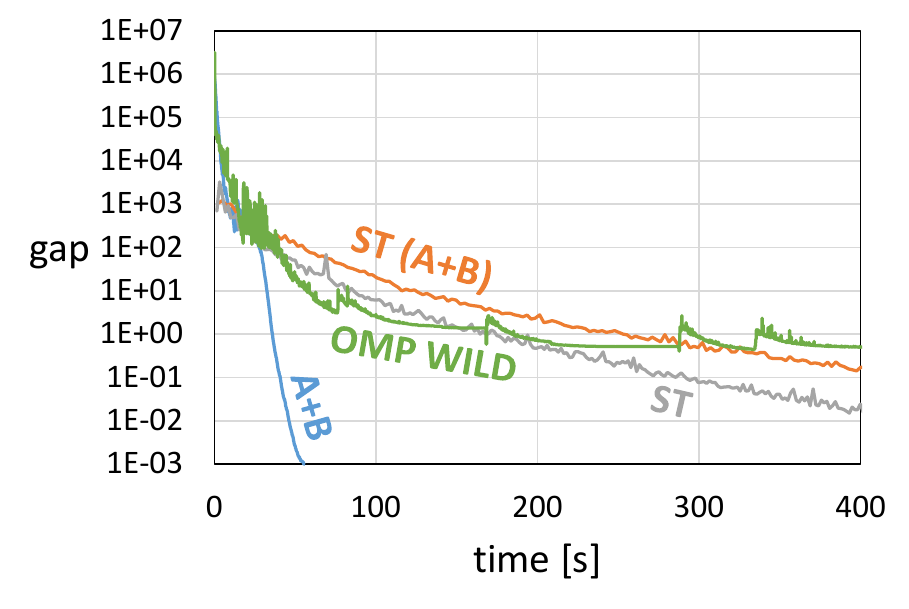}}
\subfigure{\label{fig:svm-dvc}\includegraphics[width=0.33\columnwidth]{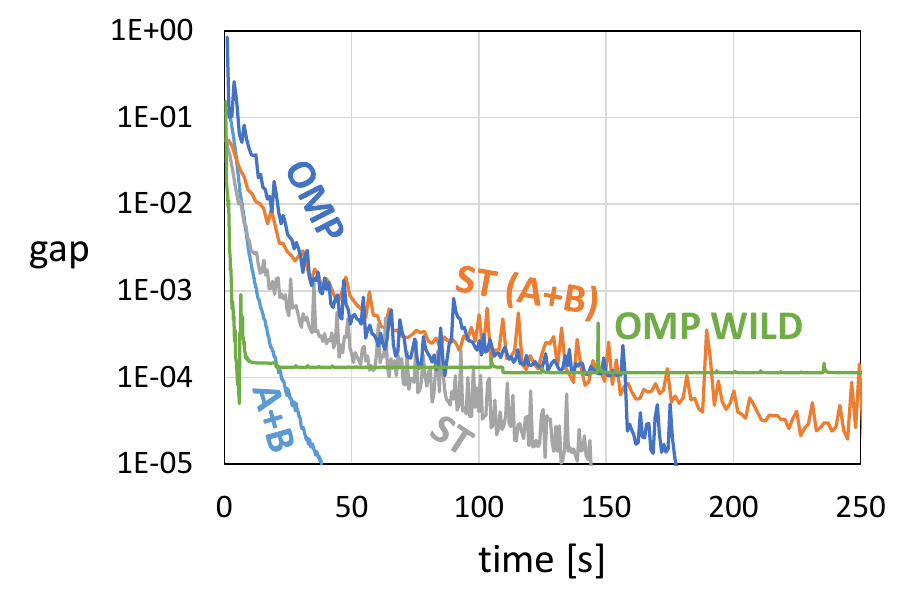}}
\end{minipage}
 \begin{minipage}[t]{\textwidth}
 \begin{tabular}{C{.39\textwidth} C{.22\textwidth} C{.4\textwidth}}
  Lasso, News20 & Lasso, News20 & SVM, News20 \\
  \end{tabular}
\subfigure{\label{fig:lasso-news-sub}\includegraphics[width=0.33\columnwidth]{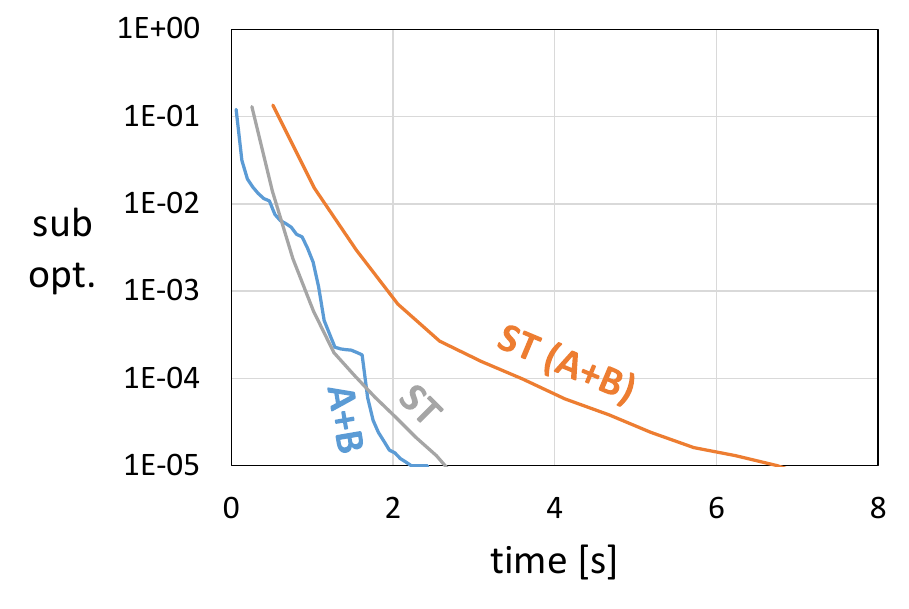}}
\subfigure{\label{fig:lasso-news}\includegraphics[width=0.33\columnwidth]{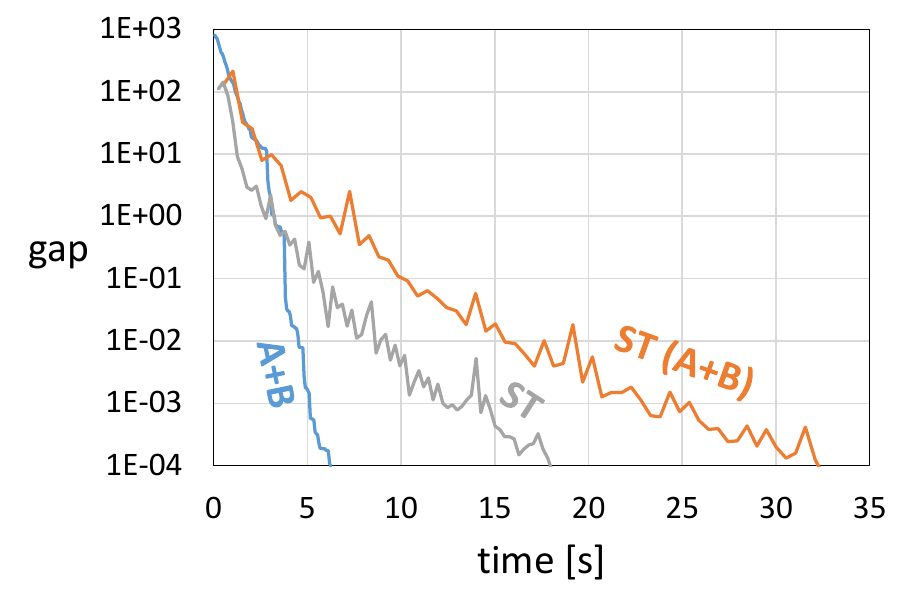}}
\subfigure{\label{fig:svm-news}\includegraphics[width=0.33\columnwidth]{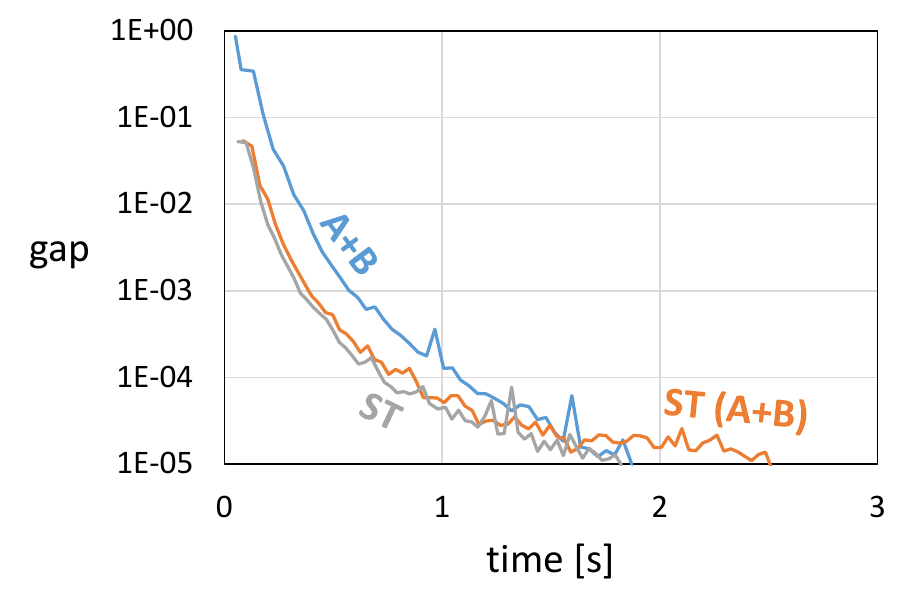}}
\end{minipage}
 \begin{minipage}[t]{\textwidth}
 \begin{tabular}{C{.39\textwidth} C{.22\textwidth} C{.4\textwidth}}
  Lasso, Criteo & Lasso, Criteo & SVM, Criteo \\
  \end{tabular}
\subfigure{\label{fig:lasso-criteo-sub}\includegraphics[width=0.33\columnwidth]{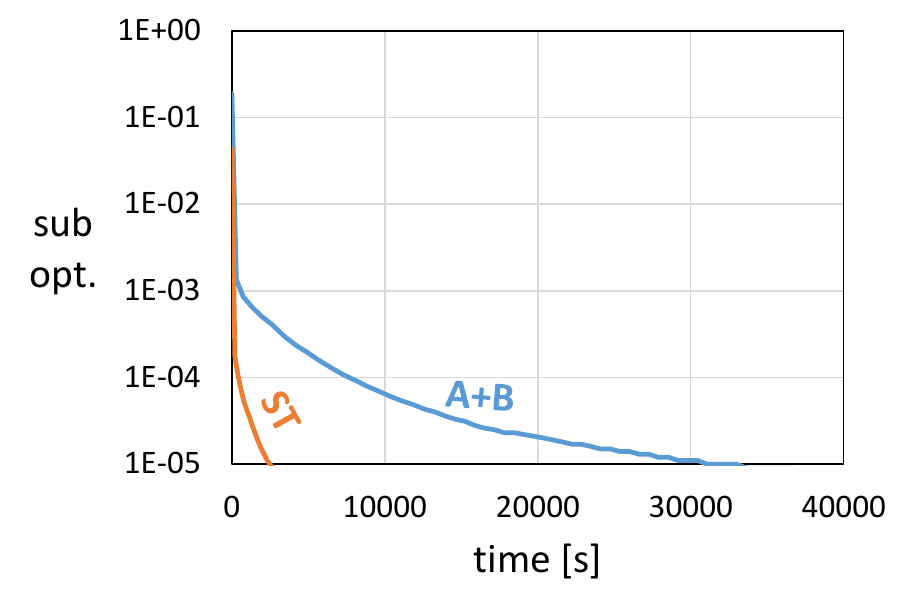}}
\subfigure{\label{fig:lasso-criteo}\includegraphics[width=0.33\columnwidth]{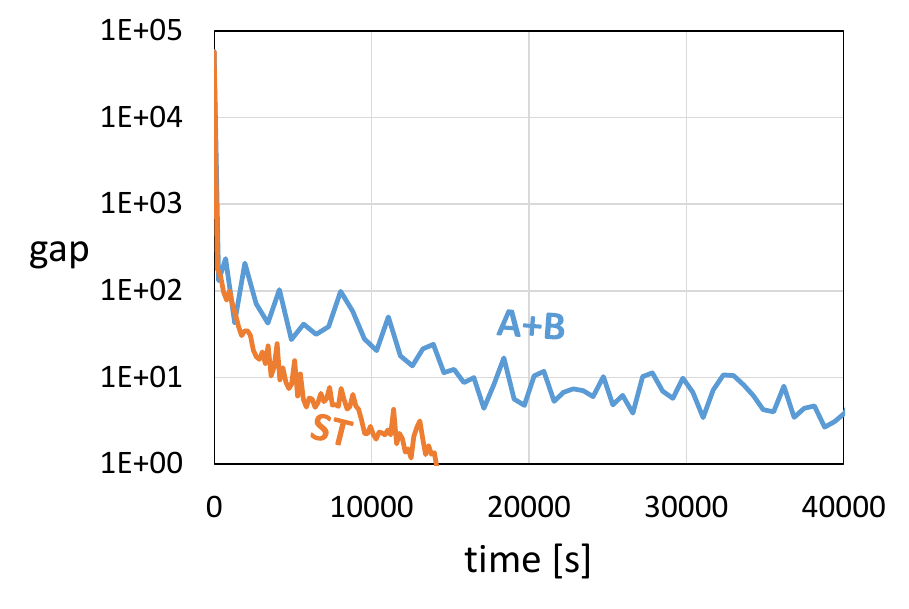}}
\subfigure{\label{fig:svm-criteo}\includegraphics[width=0.33\columnwidth]{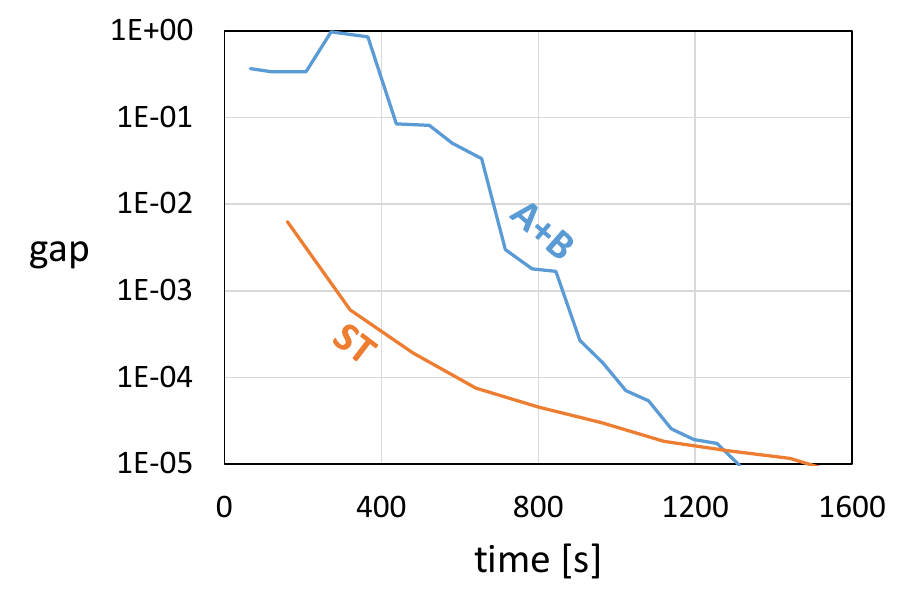}}
\caption{Convergence for Epsilon, Dogs vs. Cats, News20 and Criteo.}
\label{fig:convergence}
\end{minipage}
\end{figure*}

Fig.~\ref{fig:convergence} shows the results for Lasso and SVM for each data set. Each plot shows the precision of the algorithm versus the running time.
For Lasso, we measure suboptimality, 
for Lasso and SVM we show the duality gap.\footnote{To compute the duality gap for Lasso we use the Lipschitzing trick from~\citet{primaldualcertificates}.}
Each algorithm is run until the duality gap reaches a parametrized threshold value or until timeout. 

First, we discuss $\A+\B$, ST, and ST ($\A+\B$). For all Lasso runs, we observe a speedup varying from about 5$\times$ for Epsilon to about 9$\times$ for News20 compared to the best ST run, depending on the desired precision. As expected, ST ($\A+\B$) is never better than ST since the latter uses the best parameters found during the search. The results for suboptimality are consistent with those for the duality gaps. 

For the SVM runs, we achieve 3.5$\times$ speedup for Dogs vs. Cats and competitive performance for Epsilon and News20.

For Criteo we observe that the ST implementations beats $\A+\B$. This is mostly due to skipping the update $\vv = \vv + \dv_i \times \delta$ when $\delta = 0$. This leads to selection of relevant data based on the result of $\langle\vv, \dv_i \rangle$, and avoids expensive locking at the same time: thus, in some cases, ST drops enough operations to beat the execution time and the overhead of $\A+\B$.


Next we discuss the OpenMP runs. For OMP, as expected, the atomic operations severely impact performance and thus OMP WILD is much faster than OMP. While OMP WILD is also faster than the standard HTHC implementations, it does not guarantee the primal-dual relationship between ${\bf w}$ and $\alphav$ and thus does not converge to the exact minimizer; 
hence the plateau in the figures presenting suboptimality. The duality gap computation $\text{gap}_i(\alpha_i; \nwv)$ is based on $\nvv \neq D \alphav$, and thus do not correspond to the true values: therefore, the gap of OMP WILD eventually becomes smaller than suboptimality.
The OMP run fails to converge on the Dogs vs. Cats dataset with the used parameters.

\subsection{Comparison against other parallel CD implementations}

The work \cite{NIPS2017_7013} implements a similar scheme for parallelizing SCD on a heterogeneous platform: an 8-core Intel Xeon E5 x86 CPU with NVIDIA Quadro M4000 GPU accelerator (we note that this is a relatively old GPU generation: the newer accelerators would give better results). It provides results for Dogs vs. Cats with $\B$ updates set to 25\% (the largest size that fits into GPU RAM): a suboptimality of $10^{-5}$ is reached in 40 seconds for Lasso and a duality gap of $10^{-5}$ is reached in about 100 seconds for SVM. With the same percentage of $\B$ updates, HTHC needs 29 and 84 seconds, respectively. With our best setting (Fig.~\ref{fig:lasso-dvc-sub}--\ref{fig:svm-dvc}) this is reduced to 20 and 41 seconds, respectively. In summary, on this data, our solution on the standalone KNL is competitive with a state-of-the-art solution using a GPU accelerator with many more cores. We also show that its performance can be greatly improved with proper number of updates on $\B$.

Additionally, we compare the SVM runs of HTHC ($\A+\B$) and our parallel baseline (ST) against PASSCoDe~\cite{hsieh2015passcode}, a state-of-the-art parallel CD algorithm, which, however does not support Lasso. We compare SVM against the variant with atomic locks on $\vv$ (PASSCoDe-atomic) and a lock-free implementation (PASSCoDe-wild) which is faster, but does not maintain the relationship between model parameters as discussed in Section~\ref{sec:atomic}. The results are presented in Table~\ref{tab:passcode}. On Epsilon, the time to reach 85\% accuracy with 2 threads (the same as $T_B$ for ST) is 8.6 s for PASSCoDe-atomic and 3.21 s for PASSCoDe-wild, but these times decrease to 0.70 s with 24 threads and 0.64 s with 12 threads respectively. For Dogs vs. Cats, greatly increasing or decreasing the $T_B$ compared to ST did not improve the result. For Dogs vs. Cats, we are 2.4--5$\times$ faster, depending on the versions we compare. For Epsilon, we are roughly 2$\times$ faster, but exploiting the HTHC design is required to prevent slowdown.  On the other hand, PASSCoDe works about 7$\times$ faster for the News20 dataset. We attribute this to our locking scheme for $\vv$ updates, which is beneficial for dense data, but can be wasteful for sparse representations. Disabling the locks brings the ST execution time down to 0.02 s.

We also compare the Lasso runs against Vowpal Wabbit (VW)~\cite{langford2011vowpal}, which is considered a state-of-the-art library.
Since VW does not implement coordinate descent, we opt for stochastic gradient descent. We run the computation on previously cached data. We find that too many nodes cause divergence for dense datasets and opt for 10 nodes as a safe value. For News20, we use a single node. We compare the average squared error of HTHC against the progressive validation error of VW. The results are presented in Table~\ref{tab:vw}. Again we observe that while we perform well for dense data, the training on sparse data is slow. Also, the runs on our code and Vowpal Wabbit's SGD result in two different scores for News20.

Overall, we observe that our approach is more efficient for dense representations. On sparse datasets, the amount of computation is too small to compensate for synchronization overhead.

\begin{table}[t!]
\footnotesize
  \caption{Comparison of $\A+\B$ and ST against PASSCoDe (no support for Lasso) for SVM.}
  \label{tab:passcode}
  \centering
  \begin{tabular}{@{}lrrrrr@{}}
    \toprule
    Data set & Accuracy & $\A+\B$ & ST & PASSCoDe- & PASSCoDe- \\
    & & & & atomic & wild \\
    \midrule
    Epsilon & 85\% & 0.35 s & 1.11 s & 0.70 s & 0.64 s \\
    DvsC & 95\% & 0.51 s & 0.69 s & 2.69 s & 1.66 s \\
    News20 & 99\% & 0.14 s & 0.06 s & 0.02 s & 0.01 s \\
    \bottomrule
  \end{tabular}
%
\bigskip
\footnotesize
  \caption{Comparison of $\A+\B$ and ST against Vowpal Wabbit for Lasso.}
  \label{tab:vw}
  \centering
  \begin{tabular}{@{}lrrrr@{}}
  	\toprule
    Data set & Squared error & $\A+\B$ & ST & Vowpal Wabbit \\
    \midrule
    Epsilon & 0.47 & 0.56 s & 0.62 s & 12.19 s \\
    DvsC & 0.15 & 5.91 s & 23.37 s & 47.29 s \\
    News20 & 0.32 & 0.94 s & 0.76 s & 0.02 s \\
    \bottomrule
  \end{tabular}
\end{table}

\begin{figure*}[t]
\centering    
 \begin{minipage}[t]{\textwidth}
\subfigure[Lasso on Epsilon]{\label{fig:lasso-eps-sensitivity}\includegraphics[width=0.33\columnwidth]{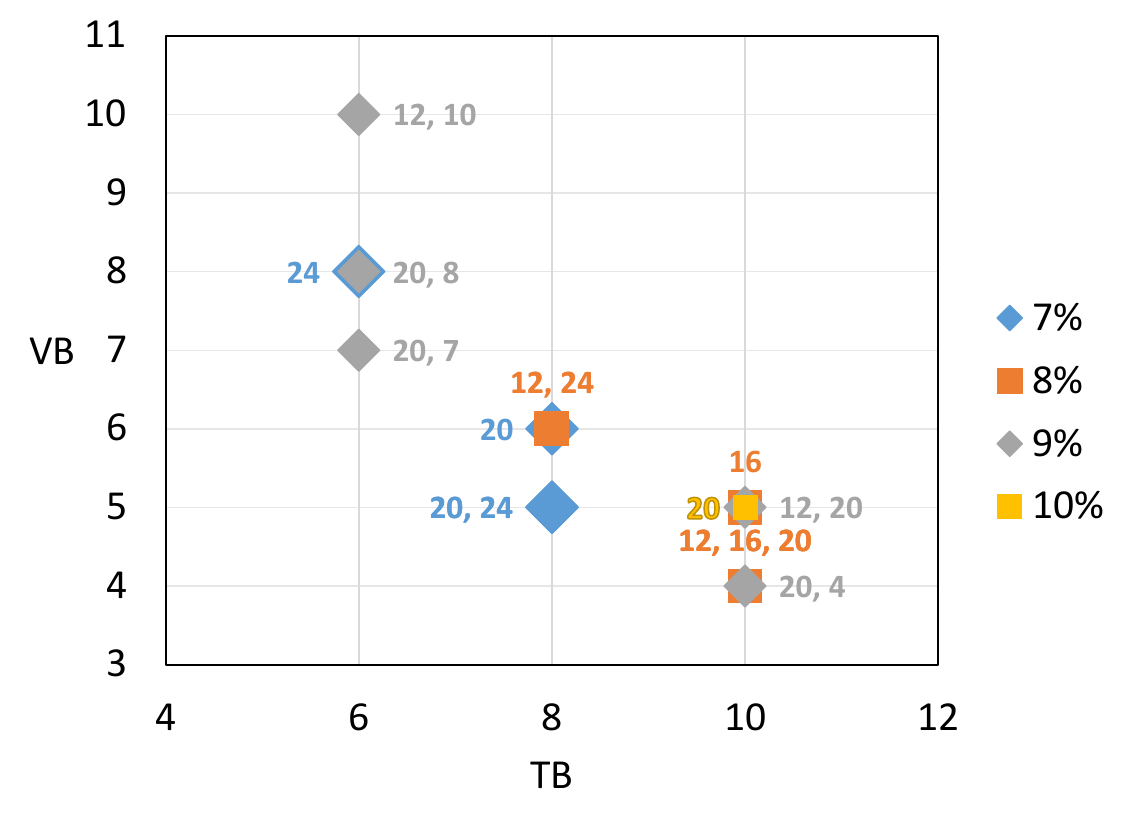}}
\subfigure[Lasso on Dogs vs. Cats]{\label{fig:lasso-dvc-sensitivity}\includegraphics[width=0.33\columnwidth]{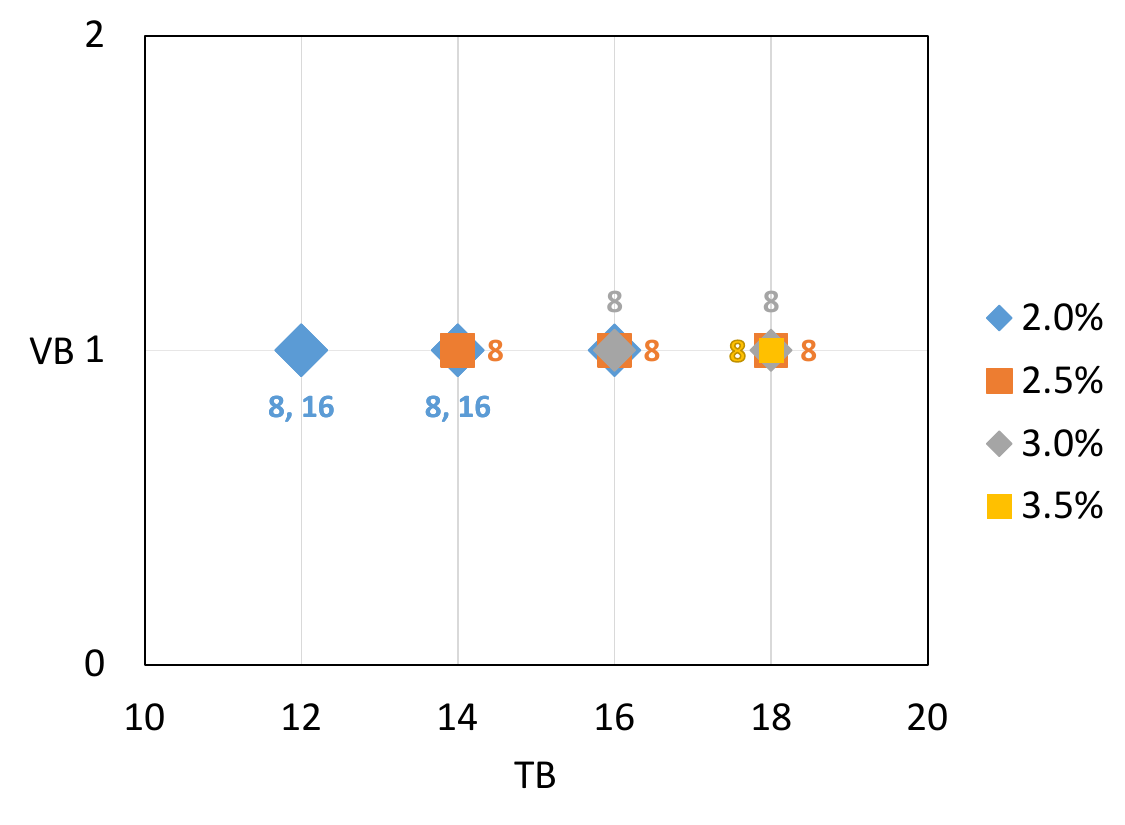}}
\subfigure[SVM on Dogs vs. Cats]{\label{fig:svm-dvc-sensitivity}\includegraphics[width=0.33\columnwidth]{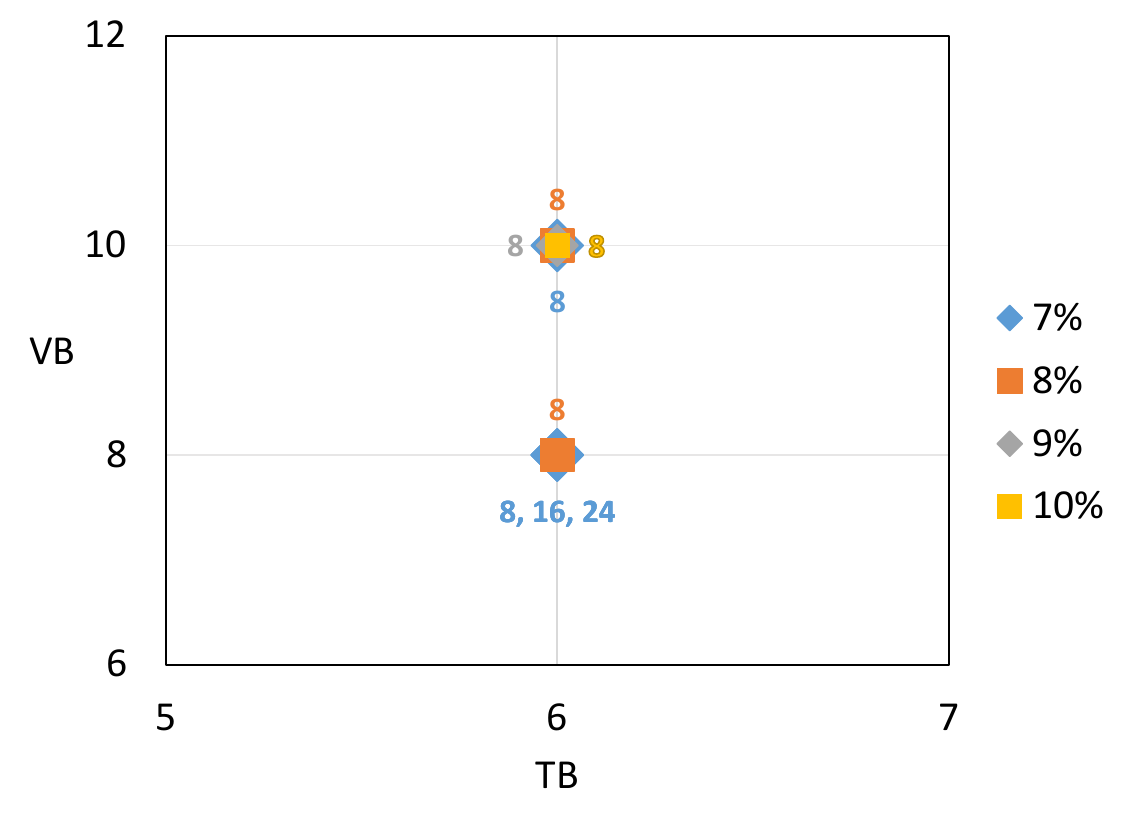}}
\caption{Parameter combinations ($T_\B$, $V_\B$) providing fast convergence (within 110\% time of the best found).}
\label{fig:sensitivity}
\end{minipage}
\end{figure*}

\begin{figure*}[tb]
\centering    
\subfigure{\includegraphics[width=\columnwidth]{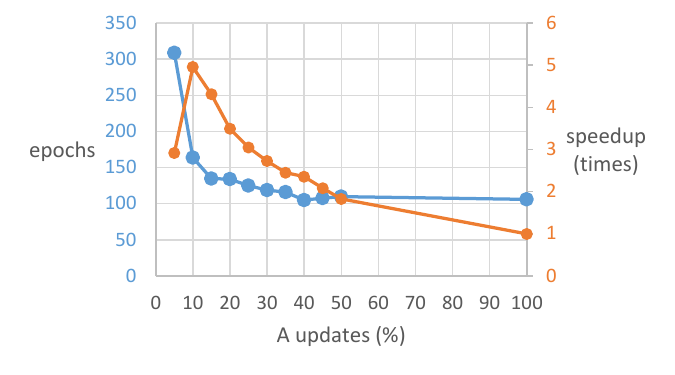}}
\subfigure{\includegraphics[width=\columnwidth]{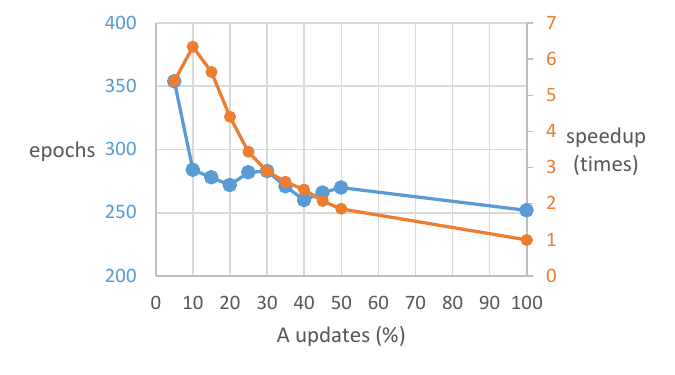}}
\caption{Sensitivity to the number of $\A$ updates per epoch for (left) Lasso on Epsilon and (right) SVM on Dogs vs. Cats.}
\label{fig:sensitivity-a}
\end{figure*}

\subsection{Experiments on sensitivity}

During the search for $\A+\B$, we considered four parameters: size of $\B$, $T_\A$, $T_\B$ and $V_\B$. Our goal was not only to find the best parameters, but also to identify parameters giving a close-to-best solution. Fig.~\ref{fig:sensitivity} presents parameters which provided no more than overall 110\% convergence time of the best solution found.  The overall convergence time depends on the number of epochs which varies from run to run for the same parameters: therefore, we consider all the presented parameters capable of obtaining the minimum runtime. The plots present four dimensions: the axes correspond to $T_\B$ and $V_\B$ while different markers correspond to different percentages of data updated by $\B$ per epoch. The labels next to the markers correspond to $T_\A$. The color of each label corresponds to its marker. Multiple values per label are possible. To save time during the search, we used a step of 8 and 4 for $T_\A$ on Dogs vs. Cats and Epsilon respectively. Additionally, we use a step of 2 for $T_\B$ on both datasets. We also note that while Lasso on Epsilon converges fast for $T_\B$ greater than 8, the rate of diverging runs is too high to consider it for practical applications.

To examine how the number of $z_i$ updates per epoch on $\A$ affects the convergence, we run tests in which we always let $\A$ perform a fixed number of updates. We use the best parameters found for different datasets and models, but we set $T_{\A} = 10$.
We present example results in Fig.~\ref{fig:sensitivity-a}.
We observe that relatively few updates are needed for the best execution time: we observe 10\% on both datasets. While these runs need more epochs to converge, the epochs are executed fast enough to provide an overall optimal convergence speed.

\begin{table}[t!]
	\footnotesize
	\caption{Comparison of 32-bit to mixed 32/4-bit.}
	\label{tab:quantized}
	\centering
	\begin{tabular}{@{}lrrrr@{}}
		\toprule
		Dataset & Model & Target gap & 32-bit & 32/4-bit \\
		\midrule
		Epsilon & Lasso & $10^{-5}$ & 1.6 s & 1.0 s \\
		Epsilon & SVM & $10^{-5}$ & 5.5 s & 5.8 s \\
		DvsC & Lasso & $10^{-3}$ & 55.5 s & 32.4 s \\
		DvsC & SVM & $10^{-5}$ & 38.2 s & 51.6 s \\
		\bottomrule
	\end{tabular}
\end{table}

\subsection{Evaluation of quantized representation}
We run experiments on the dense datasets using the quantized 4-bit representation of the data matrix $D$ with the modified Clover library. Table~\ref{tab:quantized} shows the comparison of the fastest $\A+\B$ runs using the mixed 32/4-bit arithmetic to the fastest 32-bit $\A+\B$ runs. We can observe that while we reduce the data size, the computation times do not deviate significantly from those obtained with 32-bit representation.

\section{Related Work}\label{sec:related}

Variants of 
stochastic coordinate descent~\cite{Wright2015} have become the state-of-the-art methods for training GLMs on parallel and distributed machine learning systems.
%
Parallel coordinate descent (CD) has a long history, see e.g.~\cite{Richtarik:2015br}. Recent research has contributed to asynchronous variants such as~\cite{liu2015asynchronous} who proposed AsySCD, the first asynchronous SCD algorithm, and \cite{hsieh2015passcode} who proposed the more practical PaSSCoDe algorithm which was the first to keep the shared vector $\vv$ in memory. 
%
%


There are only few works that have studied CD on non-uniform memory systems (e.g. memory and disk). The approach most related to ours is~\cite{Chang:2011:SBM:2020408.2020517} where the authors proposed a strategy to keep informative samples in memory. However,~\cite{Chang:2011:SBM:2020408.2020517} is specific to the SVM problem and unable to generalize to the broader class of GLMs. In~\cite{Matsushima:2012:LSV:2339530.2339559} a more general setting was considered, but the proposed  random (block) coordinate selection scheme is unable to benefit from non-uniformity in the training data. 
In a single machine setting, various schemes for selecting the relevant coordinates for CD have been studied, including adaptive probabilities, e.g.~\cite{Perekrestenko:226287} or fixed importance sampling~\cite{pmlr-v37-zhaoa15}.
The selection of relevant coordinates can be based on the steepest gradient, e.g.~\cite{You:2016:APG:3157382.3157621}, 
Lipschitz constant of the gradient~\cite{Zhang:2016:ASB:2939672.2939819}, nearest neighbor~\cite{Yen:2013:IBC:2487575.2487626} or duality gap based measures \cite{NIPS2017_7013}. In this work, we build on the latter, but any adaptive selection scheme could be adopted.

Manycore machines, including KNL, are  widely used for deep learning, as standalone devices or within clusters, e.g.~\cite{Gawande2017ScalingDL, You2017100epochIT}.
%
SVM training on multicore and manycore architectures was proposed by You et al.~\citet{6877312}. The authors provide evaluation for Knights Corner (KNC) and Ivy Bridge, proving them to be competitive with GPUs. The LIBSVM library~\cite{Chang:2011:LLS:1961189.1961199} is implemented for both GPU~\cite{athanasopoulos2011gpu} and KNC~\cite{10.1007/978-3-319-73353-1_20}. All SVM implementations use the sequential minimization algorithm~\cite{sequential-minimal-optimization-a-fast-algorithm-for-training-support-vector-machines}. The library and its implementations are more fitted for kernel SVM than the linear version. For training on large-scale linear models, a multi-core extension of LIBLINEAR~\cite{Fan:2008:LLL:1390681.1442794} was proposed by Chiang et al.~\citet{Chiang:2016:PDC:2939672.2939826}. This library is tailored mainly for sparse data formats used e.g. in text classification. While~\cite{Chiang:2016:PDC:2939672.2939826, hsieh2015passcode} do not perform coordinate selection, they use techniques like shrinking benefitting from increasing sparsity of the output models. Rendle et al.~\citet{Rendle:2016:RLM:2939672.2939790} introduced coordinate descent for sparse data on distributed systems, achieving almost linear scalability: their approach can be applied to multi- and manycore. The existing open-source libraries support mainly sparse data and rarely implement CD models other than SVM or logistic regression.



\section{Conclusions}\label{sec:conclusions}

We introduced HTHC for training general linear models on standalone manycores including a complete, architecture-cognizant implementation. We support dense, sparse, and quantized 4-bit data representations.
We demonstrated that HTHC provides a significant reduction of training time as opposed to a straightforward parallel implementation of coordinate descent. In our experiments, the speedup varies from 5$\times$ to more than 10$\times$ depending on the data set and the stopping criterion. We also showed that our implementation for dense datasets is competitive against the state-of-the-art libraries and a CPU-GPU code.
An advantage of HTHC over the CPU-GPU heterogeneous learning schemes is the ability of balancing distribution of machine resources such as memory and CPU cores between different tasks, an approach inherently impossible on heterogeneous platforms. To the best of our knowledge, this is the first scheme with major heterogeneous tasks running in parallel proposed in the field of manycore machine learning. The inherent adaptivity of HTHC should enable porting it to other existing and future standalone manycore platforms. 

\bibliographystyle{ieeetran}
\bibliography{./ms}

\end{document}

%% file: duhl-diag.tex
\tikzstyle{graybox} = [draw=black, fill=gray!10, rectangle, scale=0.7]
\tikzstyle{peachbox} = [draw=black, fill=orange!15, rectangle, scale=0.7]
\tikzstyle{whitebox} = [draw=black, fill=white, rectangle, scale=0.7]
\tikzstyle{greenbox} = [draw=black, fill=green!15, rectangle, scale=0.7]
\tikzstyle{wrapper} = [draw=black, fill=gray!15, rectangle, dashed, minimum width=0.84\columnwidth, minimum height=76]
\tikzstyle{thinarrow} = [-latex, draw=black, line width=0.5]
\tikzstyle{thinline} = [draw=black, line width=0.5]

\begin{tikzpicture}[node distance = 0.2 and -1.15]

\node[graybox] (initbox){
\begin{minipage}{0.6\columnwidth}
\begin{center}
\begin{tabular}{ll}
$t = 0$: & initialize $\alphav, \vv$ \\
 & set $\zv = 0$
\end{tabular}
\end{center}
\end{minipage}
};

\node[wrapper, below = 0.46 of initbox] (epoch) {};

\node[graybox, below = 0.54 of initbox] (findbox){
\begin{minipage}{1.12\columnwidth}
\begin{center}
Determine set $\cP$ of coordinates:
\vspace{-1mm}
\begin{align*}
\cP = \argmax_{\cP \subset \left[ n \right] : | \cP | = m} \sum_{i \in \cP} z_i
\end{align*}
\end{center}
\end{minipage}
};

\node[greenbox, below left = 0.1 and -0.3975\columnwidth of findbox] (abox){
\begin{minipage}{0.54\columnwidth}
\textbf{Task $\bm{\A}$} \\
\textit{while \textbf{Task $\bm{\B}$} not finished:} \\
\phantom{abc} randomly sample $i \in \left[ n \right]$ \\
\phantom{abc} $z_i = \text{gap}_i(\alpha_i^t; w(\vv^t))$
\vspace{1mm}
\end{minipage}
};

\node[peachbox, below right = 0.1 and -0.3975\columnwidth of findbox] (bbox){
\begin{minipage}{0.54\columnwidth}
\textbf{Task $\bm{\B}$} \\
optimize on $\{\dv_i\}_{i \in \cP}$ and \\
update model vector \\
\phantom{a} $(\alphav^{t}, \vv^{t}) \rightarrow (\alphav^{t+1}, \vv^{t+1})$
\vspace{1mm}
\end{minipage}
};

\draw[thinarrow] (initbox.south) -- (epoch.north);

\coordinate[below = 0.23 of epoch] (c1) {};
\coordinate[below right = 0.23 and 0.6 of epoch.south east] (c2) {};
\coordinate[above right = 0.23 and 0.6 of epoch.north east] (c3) {};
\coordinate[above = 0.23 of epoch] (c4) {};

\draw[thinline] (epoch.south) -- (c1);
\draw[thinline] (c1) -- (c2);
\draw[thinline] (c2) -- (c3);
\draw[thinarrow] (c3) -- (c4);

\node[graybox, right = 0.07 of epoch] (inct) {$t \leftarrow t + 1$};


\end{tikzpicture}